\documentclass[aps,prl,twocolumn,showpacs]{revtex4}
\usepackage{graphicx}
\usepackage{textcomp}
\begin{document}

\title{\emph{Ab initio} Study of Graphene on
  SiC}

\author{Alexander Mattausch}
\email{Alexander.Mattausch@physik.uni-erlangen.de}

\author{Oleg Pankratov}

\affiliation{Theoretische Festk\"orperphysik, Universit\"at
  Erlangen-N\"urnberg, Staudtstr. 7, 91058 Erlangen, Germany}

\date{\today}

\begin{abstract}
  Employing density-functional calculations we study single and double
  graphene layers on Si- and C-terminated $1 \times 1$ - 6H-SiC surfaces. We
  show that, in contrast to earlier assumptions, the first carbon layer is
  covalently bonded to the substrate, and cannot be responsible for the
  graphene-type electronic spectrum observed experimentally. The
  characteristic spectrum of free-standing graphene appears with the second
  carbon layer, which exhibits a weak van der Waals bonding to the underlying
  structure. For Si-terminated substrate, the interface is metallic, whereas
  on C-face it is semiconducting or semimetallic for single or double graphene
  coverage, respectively.
\end{abstract}

\pacs{68.35.Ct, 68.47.Fg, 73.20.-r}

\bibliographystyle{apsrev}

\maketitle

The last years have witnessed an explosion of interest in the prospect of
graphene-based nanometer-scale electronics~\cite{Be06,Ha06,Se06a,Zh06}.
Graphene, a single hexagonally ordered layer of carbon atoms, has a unique
electronic band structure with the conic ``Dirac points'' at two inequivalent
corners of the two-dimensional Brillouin zone. The electron mobility may be
very high and lateral patterning with standard lithography methods allows
device fabrication~\cite{Be06}. Two ways of obtaining graphene samples have
been used up to now. In the first ``mechanical'' method, the carbon monolayers
are mechanically split off the bulk graphite crystals and deposited onto a
SiO$_{2}$/Si substrate~\cite{Zh06}. This way an almost ``free-standing''
graphene is produced, since the carbon monolayer is practically not coupled to
the substrate. The second method uses epitaxial growth of graphite on
single-crystal silicon carbide (SiC). The ultrathin graphite layer is formed
by vacuum graphitization due to Si depletion of the SiC surface~\cite{Fo98}.
This method has apparent technological advantages over the ``mechanical''
method, however it does not guarantee that an ultrathin graphite (or graphene)
layer is electronically isolated from the substrate. Moreover, one expects a
covalent coupling between both which may strongly modify the electronic
properties of the graphene overlayer. Yet, experiments show that the transport
properties of the interface are dominated by a single epitaxial graphene
layer~\cite{Be06,Ha06}. Most surprisingly, the electronic spectrum seems not
to be affected much by the substrate. As in free-standing graphene one
observes the ``Dirac points'' with the linear dispersion relation around them.
The electron dynamics is governed by a Dirac-Weyl Hamiltonian with the Fermi
velocity of graphene replacing the speed of light. This leads to an unusual
sequence of Landau levels in a magnetic field and hence to peculiar features
in the quantum Hall effect~\cite{Be06,Zh06}.

The growth of high-quality graphene layers on both Si-terminated or
C-terminated SiC\{$0001$\} surfaces occurs in vacuum at annealing temperatures
above 1400$^{\circ}$C. The geometric structure of the interface is unclear.
Forbeaux \emph{et al.}~\cite{Fo98} proposed that on the Si-face the graphite
layer is loosely bound by van der Waals-forces to the $\sqrt{3} \times
\sqrt{3}R30^{\circ}$-reconstructed substrate. On the contrary, combining STM
and LEED data with DFT calculations Chen \emph{et al.}~\cite{Ch05} came to the
conclusion that the graphite sheet is formed on a complex $6 \times
6$-structure, from which originates the observed $6\sqrt{3} \times
6\sqrt{3}R30^{\circ}$ reconstruction that precedes the graphite formation. On
the C-terminated SiC($000\bar{1}$) face, graphite growth on top of a $2\times
2$ reconstruction was reported~\cite{Fo98,Fo00}. Berger \emph{et
  al.}~\cite{Be06,Ha06} observed the formation of large high-quality graphene
islands on top of a $1\times 1$ C-terminated SiC substrate with a
$\sqrt{3}\times \sqrt{3}R30^{\circ}$ interface reconstruction.

In this work we employ an \emph{ab initio} density-functional theory approach
to study the bonding and electronic structure of graphene on SiC. We find that
a strong covalent bonding of the first carbon layer to the substrate removes
the graphene-type electronic features from the energy region around the Fermi
level. However, these features reappear with the second carbon layer. We also
compare the electronic properties of graphene on Si- and C-terminated
surfaces.

\begin{figure}
  \includegraphics[width=\linewidth]{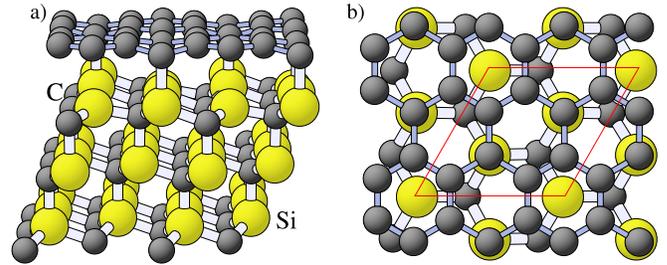}

  \caption{\label{fig:view}(Color online) Side view (a) and top view (b) of a
    graphene layer on the SiC(0001) surface. The $\sqrt{3} \times
    \sqrt{3}R30^{\circ}$ surface unit cell is highlighted.}
\end{figure}

Our calculations were performed with the density-functional theory program
package VASP~\cite{Kr93,Kr93th,Kr96,Kr96a} in the local spin density
approximation (LSDA). Projector augmented wave (PAW)
pseudopotentials~\cite{Kr99} were used. A special $7\times 7\times 1$
$\mathbf{k}$-point sampling was applied for the Brillouin-zone integration.
The plane wave basis set was restricted by a cut-off energy of 400\,eV. We
have chosen a 6H-SiC polytype, which is most often used in experimental
studies. The supercell was constructed of 6 bi-layers of SiC in the
S3-structure~\cite{St99}, one or two carbon monolayers and a vacuum interval
needed to separate the slabs. The vacuum separation varied, depending on the
carbon coverage, between 10 to 15\,\AA. The graphene layer was placed on top
of the unreconstructed 6H-SiC substrate such that the structure had a lateral
$\sqrt{3}\times \sqrt{3}R30^{\circ }$ elementary cell (Fig.~\ref{fig:view}a).
Due to the lattice mismatch of 8\% between SiC and graphite, this requires
stretching the graphene layer. We verified that for the free-standing graphene
layer the stretch reduces the total bandwidth from 19.1\,eV to 17.3\,eV but
does not affect the electronic spectrum close to the Fermi energy. The elastic
energy is 0.8\,eV per graphene unit cell.

The interface unit cell (cf.\ Fig.~\ref{fig:view}b) contains three surface
atoms of the substrate and four elementary unit cells of graphene. The
dangling bonds of the substrate atoms at the corners of the unit cell are
unsaturated, while the other surface atoms bind to two carbon atoms of the
hexagonal graphene ring. In case of the Si-terminated SiC($0001$) surface, we
find that the graphene layer is separated by 2.58\,\AA\ from the SiC
substrate. The carbon atoms covalently bonded to the substrate relax towards
the SiC surface, such that the bond length is 2.0\,\AA. This is only slightly
longer than the bond length 1.87\,\AA\ in SiC. The graphene bonding releases
0.72\,eV per graphene unit cell. For the C-terminated SiC($000\bar{1}$) face,
the graphene layer is somewhat closer (2.44\,\AA) to the substrate\ and the
bond length between the bonding carbon atoms reduces to 1.87\,\AA. The energy
gain is 0.60\,eV per graphene unit cell. On both interfaces, the bonding atom
of the substrate relaxes outwards, whereas the partner graphene atom moves
towards the substrate. The bonding energies are quite close but somewhat
smaller than the elastic deformation energy of the graphene layer. However,
the latter can be drastically lessened by defects which result from the
lattice mismatch.

For a second graphene layer placed in the graphite-type AB stacking, we find a
weak bonding at a distance of 3.3\,\AA, very close to the bulk graphite value
3.35\,\AA. This conforms to the fact that LSDA, despite the lack of long-range
non-local correlations, produces reasonable interlayer distances in van der
Waals crystals like graphite~\cite{Ch94,Oo06} or $h$-BN~\cite{MaGa06}. As
shown by Marini \emph{et al.}~\cite{MaGa06}, a delicate error cancellation
between exchange and correlation underlies this apparent performance of the
LSDA. The semilocal GGA, which violates this balance, fails to generate the
interplanar bonding in both graphite~\cite{Oo06} and $h$-BN~\cite{MaGa06},
while producing a band structure identical to LSDA~\cite{Oo06}. It is thus
natural to assume that in our situation the bonding between the graphene
layers is the same as in bulk graphite with the same interplanar distance. To
reduce the calculational cost, we fixed the interplanar distance at this
value.

The first graphene layer, which is covalently bonded to the substrate, thus
serves as a buffer separating the SiC crystal and the van der Waals bonded
second graphene sheet. Most probably, the $6\sqrt{3}\times
6\sqrt{3}R30^{\circ}$ reconstructed carbon-rich Si-terminated surface observed
as a precursor of graphitization is a natural realization of this buffer layer
in the epitaxial process. The $6\sqrt{3}\times 6\sqrt{3}$ structure is
practically commensurate with graphene since 13 times the graphene lattice
constant almost precisely fits $6\sqrt{3}$ times the SiC lattice parameter. In
any case, there is no stress in the second carbon layer. Even placed on a
strongly stretched buffer layer, the upper layer relaxes to its natural
lattice constant due to the weak interlayer interaction. For the C-face Berger
\emph{et al.}~\cite{Be06} found graphene formation on a $1\times 1$ substrate
with a $\sqrt{3}\times \sqrt{3}$ interface unit cell. This structure is the
same as we used in our calculations.

\begin{figure*}[t]
  \begin{center}
    \includegraphics[width=0.8\linewidth]{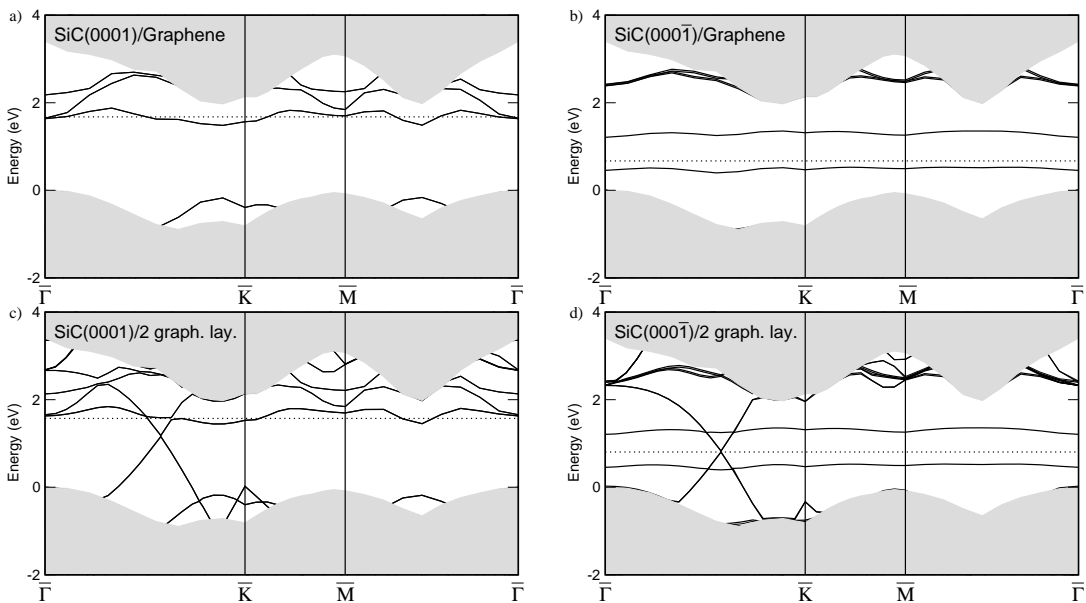}
  \end{center}
  \caption{\label{fig:bandstructure} Energy spectrum of the interface states
    of a) the SiC($0001$)/graphene interface, b) the
    SiC($000\bar{1}$)/graphene interface, c) SiC($0001$) with two layers of
    graphene and d) SiC($000\bar{1}$) with two layers of graphene. The Fermi
    energy is indicated by the dashed line. $\bar{K}$ and $\bar{M}$ are the
    high-symmetry points of the surface Brillouin zone of the $\sqrt{3} \times
    \sqrt{3}R30^{\circ}$ surface unit cell.}
  
\end{figure*}

Figs.~\ref{fig:bandstructure}a and~\ref{fig:bandstructure}b show the
electronic energy spectrum of a single graphene layer on the two SiC surfaces.
The shaded regions are the projected conduction and valence energy bands of
SiC. The Kohn-Sham energy gap of 1.98\,eV is smaller than the optical band gap
(3.02\,eV) of the bulk 6H-SiC, which is a common consequence of LSDA. The
covalent bonding drastically changes the graphene electron spectrum at the
Fermi energy. The ``Dirac cones'' are merged into the valence band, whereas
the upper graphene bands overlap with the SiC conduction band. Hence a wide
energy gap emerges in the graphene spectrum. A similar gap opening due to
hydrogen absorption on a single graphene sheet was predicted in
Ref.~\onlinecite{Du04}. The weakly dispersive interface states visible in
Figs.~\ref{fig:bandstructure}a and~~\ref{fig:bandstructure}b result from the
interaction of the graphene layer with the three dangling orbitals of the
substrate. Two of them make covalent bonds, while the third one in the center
of the graphene ring remains unsaturated (cf.\ Fig.~\ref{fig:view}b). A
projection analysis of the wave functions reveals that the gap states close to
the Fermi energy originate from the remaining dangling bonds of the substrate.
On the Si-face we find a half-filled metallic state, whereas on the C-face the
interface state is split into a singly occupied (spin polarized) and an empty
state, making the interface insulating. In contrast, on both clean SiC
surfaces LSDA predicts a substantial splitting of the surface states (0.86\,eV
for SiC($0001$) and 0.45\,eV for SiC($000\bar{1}$), see
Table~\ref{tab:electronic-structure}). Actually, the gap separating a singly
occupied and an empty state is larger due to the Hubbard repulsion of the
electrons (about 2\,eV for the $\sqrt{3} \times \sqrt{3}R30^{\circ}$
reconstructed surface~\cite{Ro00,An00}), but already LSDA correctly reproduces
the insulating character of both surfaces.

\begin{figure}
  \begin{center}
    \includegraphics[width=\linewidth]{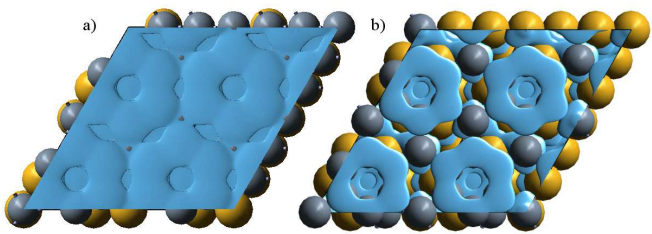}
  \end{center}
  \caption{\label{fig:charge}(Color online) Charge density of the interface
    states at the Fermi energy for a single graphene layer on a) SiC($0001$)
    and b) SiC($000\bar{1}$).}
  
\end{figure}

The reason for the striking difference between the two graphene-covered
surfaces becomes clear if one compares the planar localization of the two gap
states. As seen in Fig.~\ref{fig:charge}a for the Si-face the interface state
electron density is strongly delocalized. As the projection analysis shows,
this results from the hybridization with the graphene-induced electron states
overlapping with the conduction band (see Fig.~\ref{fig:bandstructure}a).
Given the delocalized nature of the interface state we expect the influence of
Hubbard correlations to be small. In contrast, at the C-terminated substrate
the electron state retains its localized character, although it is smeared
over a carbon ring just above the unsaturated C-dangling bond. The
localization favors the spin polarization and thus the splitting of the gap
state, whereas the interface state at the Si-face remains spin-degenerate. In
the former case, Hubbard correlations may lead to a further splitting of the
interface state.

\begin{table*}
  \caption{\label{tab:electronic-structure} Parameters of the unreconstructed
    and graphene-covered SiC$\{0001\}$ surfaces in eV: work function $\phi$,
    positions of the occupied and the unoccupied surface and interface states
    above the valence band edge ($E_{\text{o}}$, $E_{\text{u}}$) and their
    corresponding bandwidths ($B_{\text{o}}$, $B_{\text{u}}$).}
  \begin{ruledtabular}
    \begin{tabular}{lccccc}
      & Work function $\phi$ & $E_{\text{o}}$ & $B_{\text{o}}$ &
      $E_{\text{u}}$ & $B_{\text{u}}$\\
      \hline
      SiC($0001$) 1$\times$1 & $4.75$ & $E_{\text{v}} + 0.92$ & $0.45$ &
      $E_{\text{v}} + 1.78$ & $0.53$ \\
      SiC($0001$)/Graphene & $3.75$ &  $E_{\text{v}} + 1.64$ &  $0.35$ & $-$ & $-$\\
      SiC($0001$)/2 Graphene & $4.33$ & $E_{\text{v}} + 1.64$ & $0.40$ & $-$ &
      $-$\\
      SiC($000\bar{1}$) 1$\times$1 & $5.75$ & $E_{\text{v}} + 0.05$ & $0.75$ &
      $E_{\text{v}} + 0.50$ & $0.45$\\
      SiC($000\bar{1}$)/Graphene & $5.33$ & $E_{\text{v}} + 0.43$ & $0.13$ &
      $E_{\text{v}} + 1.19$ & $0.14$ \\
      SiC($000\bar{1}$)/2 Graphene &  $5.31$ & $E_{\text{v}} + 0.44$ & $0.10$
      & $E_{\text{v}} + 1.19$ & $0.15$\\
      Graphene (single layer) & $5.11$ \\
    \end{tabular}
  \end{ruledtabular}
\end{table*}

Figures~\ref{fig:bandstructure}c and \ref{fig:bandstructure}d show that the
second carbon layer indeed possesses an electronic structure similar to
free-standing graphene. The characteristic conic point appears on the
$\bar{\Gamma}-\bar{K}$ line (note that since the Brillouin zone corresponds to
the $\sqrt{3}\times \sqrt{3}R30^{\circ}$ unit cell, the conic point is not
located at the $\bar{K}$-point). The interface states of the buffer layer
remain practically unchanged since the interaction of the carbon layers is
very small. The metallic interface state on the Si-terminated substrate pins
the Fermi level just above the conic point, making the second graphene layer
$n$-doped. On C-terminated substrate the Fermi level runs exactly through the
conic point. Hence the interface is semimetallic just as for free-standing
graphene. Indeed, for a graphene-covered C-face Berger \emph{et
  al.}~\cite{Be06} found that the thin graphite layers possess electronic
properties of free-standing graphene.

The parameters of the electron states for the different interfaces are
summarized in Table~\ref{tab:electronic-structure}. For clean unreconstructed
surfaces we find work functions of 4.75\,eV (Si-terminated surface) and
5.75\,eV (C-terminated surface). The former value is practically the same as
the work function of the reconstructed SiC($0001$)~\cite{Wi03}. The first
graphene layer reduces this value to 3.75\,eV, which is 1.3\,eV lower than the
work function of free-standing graphene. The drastic reduction of the work
function is caused by charge flow from graphene to the interface region, which
induces a dipole layer. On the C-face the graphene overlayer also reduces the
work function, but to a lesser extent such that it remains above the graphene
value. Adding the second graphene layer makes the work function closer to that
of graphene for both faces.

The Fermi level pinning close to the conduction band makes the graphitized
Si-face especially suitable for Ohmic contacts on $n$-type SiC, because it
guarantees a low Schottky barrier. Indeed, Lu \emph{et al.}~\cite{Lu03} find a
very low resistance for thermally treated SiC contacts with nickel and cobalt,
while other metals, which form carbides and thereby remove the graphitic
inclusions, were rectifying. Recently Seyller \emph{et al.} measured the
Schottky barrier between $n$-type 6H-SiC($0001$) and graphite by photoelectron
spectroscopy and found a low value of 0.3\,eV~\cite{Se06}. On the contrary,
the C-terminated face has the Fermi level close to the middle of the band gap
and is semiconducting or semimetallic.

In conclusion, we investigated the interface between $1 \times 1$ -
6H-SiC\{$0001$\} surfaces and carbon layers employing \emph{ab initio}
density-functional theory. We find that graphene overlayers on SiC($0001$) and
SiC($000\bar{1}$) faces possess qualitatively different electronic structures.
While the former is metallic, the latter has semiconducting properties. The
conic points at the Fermi energy, which are specific for graphene, appear only
with the second layer. The first carbon sheet is covalently bound to the
substrate and plays the role of a transition region between a covalent SiC
crystal and a van der Waals bonded stack of graphene layers.

\begin{acknowledgements}
  This work was supported by Deutsche Forschungsgemeinschaft within the SiC
  Research Group. We are grateful to L. Magaud and F. Varchon for
  communicating to us similar results on the SiC/graphene
  system~\cite{Ol06_pc} and fruitful discussions.
\end{acknowledgements}


\begin{thebibliography}{23}
\expandafter\ifx\csname natexlab\endcsname\relax\def\natexlab#1{#1}\fi
\expandafter\ifx\csname bibnamefont\endcsname\relax
  \def\bibnamefont#1{#1}\fi
\expandafter\ifx\csname bibfnamefont\endcsname\relax
  \def\bibfnamefont#1{#1}\fi
\expandafter\ifx\csname citenamefont\endcsname\relax
  \def\citenamefont#1{#1}\fi
\expandafter\ifx\csname url\endcsname\relax
  \def\url#1{\texttt{#1}}\fi
\expandafter\ifx\csname urlprefix\endcsname\relax\def\urlprefix{URL }\fi
\providecommand{\bibinfo}[2]{#2}
\providecommand{\eprint}[2][]{\url{#2}}

\bibitem[{\citenamefont{Berger et~al.}(2006)\citenamefont{Berger, Song, Li, Wu,
  Brown, Naud, Mayou, Li, Hass, Marchenkov et~al.}}]{Be06}
\bibinfo{author}{\bibfnamefont{C.}~\bibnamefont{Berger}},
  \bibinfo{author}{\bibfnamefont{Z.}~\bibnamefont{Song}},
  \bibinfo{author}{\bibfnamefont{X.}~\bibnamefont{Li}},
  \bibinfo{author}{\bibfnamefont{X.}~\bibnamefont{Wu}},
  \bibinfo{author}{\bibfnamefont{N.}~\bibnamefont{Brown}},
  \bibinfo{author}{\bibfnamefont{C.}~\bibnamefont{Naud}},
  \bibinfo{author}{\bibfnamefont{D.}~\bibnamefont{Mayou}},
  \bibinfo{author}{\bibfnamefont{T.}~\bibnamefont{Li}},
  \bibinfo{author}{\bibfnamefont{J.}~\bibnamefont{Hass}},
  \bibinfo{author}{\bibfnamefont{A.~N.} \bibnamefont{Marchenkov}},
  \bibnamefont{et~al.}, \bibinfo{journal}{Science}
  \textbf{\bibinfo{volume}{312}}, \bibinfo{pages}{1191} (\bibinfo{year}{2006}).

\bibitem[{\citenamefont{Hass et~al.}(2006)\citenamefont{Hass, Jeffrey, Feng,
  Li, Li, Song, Berger, de~Heer, First, and Conrad}}]{Ha06}
\bibinfo{author}{\bibfnamefont{J.}~\bibnamefont{Hass}},
  \bibinfo{author}{\bibfnamefont{C.~A.} \bibnamefont{Jeffrey}},
  \bibinfo{author}{\bibfnamefont{R.}~\bibnamefont{Feng}},
  \bibinfo{author}{\bibfnamefont{T.}~\bibnamefont{Li}},
  \bibinfo{author}{\bibfnamefont{X.}~\bibnamefont{Li}},
  \bibinfo{author}{\bibfnamefont{Z.}~\bibnamefont{Song}},
  \bibinfo{author}{\bibfnamefont{C.}~\bibnamefont{Berger}},
  \bibinfo{author}{\bibfnamefont{W.~A.} \bibnamefont{de~Heer}},
  \bibinfo{author}{\bibfnamefont{P.~N.} \bibnamefont{First}}, \bibnamefont{and}
  \bibinfo{author}{\bibfnamefont{E.~H.} \bibnamefont{Conrad}},
  \bibinfo{journal}{Appl.\ Phys.\ Lett.} \textbf{\bibinfo{volume}{89}},
  \bibinfo{pages}{143106} (\bibinfo{year}{2006}), \eprint{cond-mat/0604206}.

\bibitem[{\citenamefont{Seyller
  et~al.}(2006{\natexlab{a}})\citenamefont{Seyller, Emtsev, Gao, Speck, Ley,
  Tadich, Broekmann, Riley, Leckey, Rader et~al.}}]{Se06a}
\bibinfo{author}{\bibfnamefont{T.}~\bibnamefont{Seyller}},
  \bibinfo{author}{\bibfnamefont{K.~V.} \bibnamefont{Emtsev}},
  \bibinfo{author}{\bibfnamefont{K.}~\bibnamefont{Gao}},
  \bibinfo{author}{\bibfnamefont{F.}~\bibnamefont{Speck}},
  \bibinfo{author}{\bibfnamefont{L.}~\bibnamefont{Ley}},
  \bibinfo{author}{\bibfnamefont{A.}~\bibnamefont{Tadich}},
  \bibinfo{author}{\bibfnamefont{L.}~\bibnamefont{Broekmann}},
  \bibinfo{author}{\bibfnamefont{J.~D.} \bibnamefont{Riley}},
  \bibinfo{author}{\bibfnamefont{R.~C.~G.} \bibnamefont{Leckey}},
  \bibinfo{author}{\bibfnamefont{O.}~\bibnamefont{Rader}},
  \bibnamefont{et~al.}, \bibinfo{journal}{Surf.\ Sci.}
  \textbf{\bibinfo{volume}{600}}, \bibinfo{pages}{3906}
  (\bibinfo{year}{2006}{\natexlab{a}}).

\bibitem[{\citenamefont{Zhang et~al.}(2006)\citenamefont{Zhang, Jiang, Small,
  Purewal, Tan, Fazlollahi, Chudow, Jaszczak, Stormer, and Kim}}]{Zh06}
\bibinfo{author}{\bibfnamefont{Y.}~\bibnamefont{Zhang}},
  \bibinfo{author}{\bibfnamefont{Z.}~\bibnamefont{Jiang}},
  \bibinfo{author}{\bibfnamefont{J.~P.} \bibnamefont{Small}},
  \bibinfo{author}{\bibfnamefont{M.~S.} \bibnamefont{Purewal}},
  \bibinfo{author}{\bibfnamefont{Y.-W.} \bibnamefont{Tan}},
  \bibinfo{author}{\bibfnamefont{M.}~\bibnamefont{Fazlollahi}},
  \bibinfo{author}{\bibfnamefont{J.~D.} \bibnamefont{Chudow}},
  \bibinfo{author}{\bibfnamefont{J.~A.} \bibnamefont{Jaszczak}},
  \bibinfo{author}{\bibfnamefont{H.~L.} \bibnamefont{Stormer}},
  \bibnamefont{and} \bibinfo{author}{\bibfnamefont{P.}~\bibnamefont{Kim}},
  \bibinfo{journal}{Phys.\ Rev.\ Lett.} \textbf{\bibinfo{volume}{96}},
  \bibinfo{pages}{136806} (\bibinfo{year}{2006}).

\bibitem[{\citenamefont{Forbeaux et~al.}(1998)\citenamefont{Forbeaux, Themlin,
  and Debever}}]{Fo98}
\bibinfo{author}{\bibfnamefont{I.}~\bibnamefont{Forbeaux}},
  \bibinfo{author}{\bibfnamefont{J.-M.} \bibnamefont{Themlin}},
  \bibnamefont{and} \bibinfo{author}{\bibfnamefont{J.-M.}
  \bibnamefont{Debever}}, \bibinfo{journal}{Phys.\ Rev.\ B}
  \textbf{\bibinfo{volume}{58}}, \bibinfo{pages}{16396} (\bibinfo{year}{1998}).

\bibitem[{\citenamefont{Chen et~al.}(2005)\citenamefont{Chen, Xu, Liu, Gao, Qi,
  Peng, Tan, Feng, Loh, and Wee}}]{Ch05}
\bibinfo{author}{\bibfnamefont{W.}~\bibnamefont{Chen}},
  \bibinfo{author}{\bibfnamefont{H.}~\bibnamefont{Xu}},
  \bibinfo{author}{\bibfnamefont{L.}~\bibnamefont{Liu}},
  \bibinfo{author}{\bibfnamefont{X.}~\bibnamefont{Gao}},
  \bibinfo{author}{\bibfnamefont{D.}~\bibnamefont{Qi}},
  \bibinfo{author}{\bibfnamefont{G.}~\bibnamefont{Peng}},
  \bibinfo{author}{\bibfnamefont{S.~C.} \bibnamefont{Tan}},
  \bibinfo{author}{\bibfnamefont{Y.}~\bibnamefont{Feng}},
  \bibinfo{author}{\bibfnamefont{K.~P.} \bibnamefont{Loh}}, \bibnamefont{and}
  \bibinfo{author}{\bibfnamefont{A.~T.~S.} \bibnamefont{Wee}},
  \bibinfo{journal}{Surf.\ Sci.} \textbf{\bibinfo{volume}{596}},
  \bibinfo{pages}{176} (\bibinfo{year}{2005}).

\bibitem[{\citenamefont{Forbeaux et~al.}(2000)\citenamefont{Forbeaux, Themlin,
  Charrier, Thibaudau, and Debever}}]{Fo00}
\bibinfo{author}{\bibfnamefont{I.}~\bibnamefont{Forbeaux}},
  \bibinfo{author}{\bibfnamefont{J.-M.} \bibnamefont{Themlin}},
  \bibinfo{author}{\bibfnamefont{A.}~\bibnamefont{Charrier}},
  \bibinfo{author}{\bibfnamefont{F.}~\bibnamefont{Thibaudau}},
  \bibnamefont{and} \bibinfo{author}{\bibfnamefont{J.-M.}
  \bibnamefont{Debever}}, \bibinfo{journal}{Appl.\ Surf.\ Sci.}
  \textbf{\bibinfo{volume}{162-163}}, \bibinfo{pages}{406}
  (\bibinfo{year}{2000}).

\bibitem[{\citenamefont{Kresse and Hafner}(1993)}]{Kr93}
\bibinfo{author}{\bibfnamefont{G.}~\bibnamefont{Kresse}} \bibnamefont{and}
  \bibinfo{author}{\bibfnamefont{J.}~\bibnamefont{Hafner}},
  \bibinfo{journal}{Phys.\ Rev.\ B} \textbf{\bibinfo{volume}{47}},
  \bibinfo{pages}{558} (\bibinfo{year}{1993}).

\bibitem[{\citenamefont{Kresse}(1993)}]{Kr93th}
\bibinfo{author}{\bibfnamefont{G.}~\bibnamefont{Kresse}}, Ph.D. thesis,
  \bibinfo{school}{Technische Universit\"at Wien}, \bibinfo{address}{Austria}
  (\bibinfo{year}{1993}).

\bibitem[{\citenamefont{Kresse and Furthm{\"u}ller}(1996{\natexlab{a}})}]{Kr96}
\bibinfo{author}{\bibfnamefont{G.}~\bibnamefont{Kresse}} \bibnamefont{and}
  \bibinfo{author}{\bibfnamefont{J.}~\bibnamefont{Furthm{\"u}ller}},
  \bibinfo{journal}{Phys.\ Rev.\ B} \textbf{\bibinfo{volume}{54}},
  \bibinfo{pages}{11169} (\bibinfo{year}{1996}{\natexlab{a}}).

\bibitem[{\citenamefont{Kresse and
  Furthm{\"u}ller}(1996{\natexlab{b}})}]{Kr96a}
\bibinfo{author}{\bibfnamefont{G.}~\bibnamefont{Kresse}} \bibnamefont{and}
  \bibinfo{author}{\bibfnamefont{J.}~\bibnamefont{Furthm{\"u}ller}},
  \bibinfo{journal}{Comput. Mat. Sci} \textbf{\bibinfo{volume}{6}},
  \bibinfo{pages}{15} (\bibinfo{year}{1996}{\natexlab{b}}).

\bibitem[{\citenamefont{Kresse and Joubert}(1999)}]{Kr99}
\bibinfo{author}{\bibfnamefont{G.}~\bibnamefont{Kresse}} \bibnamefont{and}
  \bibinfo{author}{\bibfnamefont{D.}~\bibnamefont{Joubert}},
  \bibinfo{journal}{Phys.\ Rev.\ B} \textbf{\bibinfo{volume}{59}},
  \bibinfo{pages}{1758} (\bibinfo{year}{1999}).

\bibitem[{\citenamefont{Starke et~al.}(1999)\citenamefont{Starke, Schardt,
  Bernhardt, Franke, and Heinz}}]{St99}
\bibinfo{author}{\bibfnamefont{U.}~\bibnamefont{Starke}},
  \bibinfo{author}{\bibfnamefont{J.}~\bibnamefont{Schardt}},
  \bibinfo{author}{\bibfnamefont{J.}~\bibnamefont{Bernhardt}},
  \bibinfo{author}{\bibfnamefont{M.}~\bibnamefont{Franke}}, \bibnamefont{and}
  \bibinfo{author}{\bibfnamefont{K.}~\bibnamefont{Heinz}},
  \bibinfo{journal}{Phys.\ Rev.\ Lett.} \textbf{\bibinfo{volume}{82}},
  \bibinfo{pages}{2107} (\bibinfo{year}{1999}).

\bibitem[{\citenamefont{Charlier et~al.}(1994)\citenamefont{Charlier, Gonze,
  and Michenaud}}]{Ch94}
\bibinfo{author}{\bibfnamefont{J.-C.} \bibnamefont{Charlier}},
  \bibinfo{author}{\bibfnamefont{X.}~\bibnamefont{Gonze}}, \bibnamefont{and}
  \bibinfo{author}{\bibfnamefont{J.~P.} \bibnamefont{Michenaud}},
  \bibinfo{journal}{Carbon} \textbf{\bibinfo{volume}{32}}, \bibinfo{pages}{289}
  (\bibinfo{year}{1994}).

\bibitem[{\citenamefont{Ooi et~al.}(2006)\citenamefont{Ooi, Rairkar, and
  Adams}}]{Oo06}
\bibinfo{author}{\bibfnamefont{N.}~\bibnamefont{Ooi}},
  \bibinfo{author}{\bibfnamefont{A.}~\bibnamefont{Rairkar}}, \bibnamefont{and}
  \bibinfo{author}{\bibfnamefont{J.~B.} \bibnamefont{Adams}},
  \bibinfo{journal}{Carbon} \textbf{\bibinfo{volume}{44}}, \bibinfo{pages}{231}
  (\bibinfo{year}{2006}).

\bibitem[{\citenamefont{Marini et~al.}(2006)\citenamefont{Marini,
  Garc\'ia-Gonz\'alez, and Rubio}}]{MaGa06}
\bibinfo{author}{\bibfnamefont{A.}~\bibnamefont{Marini}},
  \bibinfo{author}{\bibfnamefont{P.}~\bibnamefont{Garc\'ia-Gonz\'alez}},
  \bibnamefont{and} \bibinfo{author}{\bibfnamefont{A.}~\bibnamefont{Rubio}},
  \bibinfo{journal}{Phys.\ Rev.\ Lett.} \textbf{\bibinfo{volume}{96}},
  \bibinfo{pages}{136404} (\bibinfo{year}{2006}).

\bibitem[{\citenamefont{Duplock et~al.}(2004)\citenamefont{Duplock, Scheffler,
  and Lindan}}]{Du04}
\bibinfo{author}{\bibfnamefont{E.~J.} \bibnamefont{Duplock}},
  \bibinfo{author}{\bibfnamefont{M.}~\bibnamefont{Scheffler}},
  \bibnamefont{and} \bibinfo{author}{\bibfnamefont{P.~J.~D.}
  \bibnamefont{Lindan}}, \bibinfo{journal}{Phys.\ Rev.\ Lett.}
  \textbf{\bibinfo{volume}{92}}, \bibinfo{pages}{225502}
  (\bibinfo{year}{2004}).

\bibitem[{\citenamefont{Rohlfing and Pollmann}(2000)}]{Ro00}
\bibinfo{author}{\bibfnamefont{M.}~\bibnamefont{Rohlfing}} \bibnamefont{and}
  \bibinfo{author}{\bibfnamefont{J.}~\bibnamefont{Pollmann}},
  \bibinfo{journal}{Phys.\ Rev.\ Lett.} \textbf{\bibinfo{volume}{84}},
  \bibinfo{pages}{135} (\bibinfo{year}{2000}).

\bibitem[{\citenamefont{Anisimov et~al.}(2000)\citenamefont{Anisimov, Bedin,
  Korotin, Santoro, Scandolo, and Tosatti}}]{An00}
\bibinfo{author}{\bibfnamefont{V.~I.} \bibnamefont{Anisimov}},
  \bibinfo{author}{\bibfnamefont{A.~E.} \bibnamefont{Bedin}},
  \bibinfo{author}{\bibfnamefont{M.~A.} \bibnamefont{Korotin}},
  \bibinfo{author}{\bibfnamefont{G.}~\bibnamefont{Santoro}},
  \bibinfo{author}{\bibfnamefont{S.}~\bibnamefont{Scandolo}}, \bibnamefont{and}
  \bibinfo{author}{\bibfnamefont{E.}~\bibnamefont{Tosatti}},
  \bibinfo{journal}{Phys.\ Rev.\ B} \textbf{\bibinfo{volume}{61}},
  \bibinfo{pages}{1752} (\bibinfo{year}{2000}).

\bibitem[{\citenamefont{Wiets et~al.}(2003)\citenamefont{Wiets, Weinelt, and
  Fauster}}]{Wi03}
\bibinfo{author}{\bibfnamefont{M.}~\bibnamefont{Wiets}},
  \bibinfo{author}{\bibfnamefont{M.}~\bibnamefont{Weinelt}}, \bibnamefont{and}
  \bibinfo{author}{\bibfnamefont{T.}~\bibnamefont{Fauster}},
  \bibinfo{journal}{Phys.\ Rev.\ B} \textbf{\bibinfo{volume}{68}},
  \bibinfo{pages}{125321} (\bibinfo{year}{2003}).

\bibitem[{\citenamefont{Lu et~al.}(2003)\citenamefont{Lu, Mitchel, Landis,
  Crenshaw, and Collins}}]{Lu03}
\bibinfo{author}{\bibfnamefont{W.}~\bibnamefont{Lu}},
  \bibinfo{author}{\bibfnamefont{W.~C.} \bibnamefont{Mitchel}},
  \bibinfo{author}{\bibfnamefont{G.~R.} \bibnamefont{Landis}},
  \bibinfo{author}{\bibfnamefont{T.~R.} \bibnamefont{Crenshaw}},
  \bibnamefont{and} \bibinfo{author}{\bibfnamefont{W.~E.}
  \bibnamefont{Collins}}, \bibinfo{journal}{J.\ Appl.\ Phys.}
  \textbf{\bibinfo{volume}{93}}, \bibinfo{pages}{5397} (\bibinfo{year}{2003}).

\bibitem[{\citenamefont{Seyller
  et~al.}(2006{\natexlab{b}})\citenamefont{Seyller, Emtsev, Speck, Gao, and
  Ley}}]{Se06}
\bibinfo{author}{\bibfnamefont{T.}~\bibnamefont{Seyller}},
  \bibinfo{author}{\bibfnamefont{K.~V.} \bibnamefont{Emtsev}},
  \bibinfo{author}{\bibfnamefont{F.}~\bibnamefont{Speck}},
  \bibinfo{author}{\bibfnamefont{K.-Y.} \bibnamefont{Gao}}, \bibnamefont{and}
  \bibinfo{author}{\bibfnamefont{L.}~\bibnamefont{Ley}},
  \bibinfo{journal}{Appl.\ Phys.\ Lett.} \textbf{\bibinfo{volume}{88}},
  \bibinfo{pages}{242103} (\bibinfo{year}{2006}{\natexlab{b}}).

\bibitem[{\citenamefont{Varchon et~al.}()\citenamefont{Varchon, Magaud, and
  Olevano}}]{Ol06_pc}
\bibinfo{author}{\bibfnamefont{F.}~\bibnamefont{Varchon}},
  \bibinfo{author}{\bibfnamefont{L.}~\bibnamefont{Magaud}}, \bibnamefont{and}
  \bibinfo{author}{\bibfnamefont{V.}~\bibnamefont{Olevano}},
  \bibinfo{note}{private communication};
\bibinfo{author}{\bibfnamefont{F.}~\bibnamefont{Varchon}},
  \bibinfo{author}{\bibfnamefont{R.}~\bibnamefont{Feng}},
  \bibinfo{author}{\bibfnamefont{J.}~\bibnamefont{Hass}},
  \bibinfo{author}{\bibfnamefont{X.}~\bibnamefont{Li}},
  \bibinfo{author}{\bibfnamefont{B.~N.} \bibnamefont{Nguyen}},
  \bibinfo{author}{\bibfnamefont{C.}~\bibnamefont{Naud}},
  \bibinfo{author}{\bibfnamefont{P.}~\bibnamefont{Mallet}},
  \bibinfo{author}{\bibfnamefont{J.~Y.} \bibnamefont{Veuillen}},
  \bibinfo{author}{\bibfnamefont{C.}~\bibnamefont{Berger}},
  \bibinfo{author}{\bibfnamefont{E.~H.} \bibnamefont{Conrad}},
  \bibinfo{author}{\bibfnamefont{L.} \bibnamefont{Magaud}},
  \eprint{arXiv:cond-mat/0702311}.

\end{thebibliography}

\end{document}